\documentclass{elsart41}
\usepackage{graphics}
\usepackage{graphicx}
\usepackage{amssymb}
\begin{document}

\newcommand{\beq}{\begin{equation}}
\newcommand{\eeq}{\end{equation}}
\newcommand{\bea}{\begin{eqnarray}}
\newcommand{\eea}{\end{eqnarray}}

\newcommand{\rmd}{{\rm d}}
\newcommand{\rmi}{{\rm i}}

\newcommand{\bn}{{\bf n}}
\newcommand{\bd}{{\bf d}}
\newcommand{\bL}{{\bf L}}
\newcommand{\bD}{{\bf D}}
\def\a{\alpha}
\def\b{\beta}
\def\LL{{\mathcal{L}}}

\begin{frontmatter}



\title{Derivation of the generalized Non Linear Sigma Model in the presence of
the Dzyaloshinskii-Moriya interaction}
%

\author[AA]{L. Benfatto\corauthref{LB}},
\ead{lara.benfatto@roma1.infn.it}
\author[BB]{M. Silva Neto},\,
\author[BB]{V. Juricic}
\author[BB]{C. Morais Smith}

\address[AA]{SMC-INFM and Department of Physics, University of Rome ``La
  Sapienza'', Piazzale Aldo Moro 5, 00185, Rome, Italy}
\address[BB]{Institute for Theoretical Physics, University of Utrecht,
  P.O. Box 80.195, 3508 TD, Utrecht, The Netherlands}

\corauth[LB]{Corresponding author. Tel: +39-06-49694212
fax: +39-06-4957697}

\begin{abstract}

We derive the long-wavelength non-linear sigma model for a two-dimensional
Heisenberg system in the presence of the Dzyaloshinskii-Moriya and
pseudodipolar interactions.
We show that the
system is a non-conventional easy-axis antiferromagnet, displaying an
anomalous coupling between the magnetic field and the staggered order
parameter. Our results are in good agreement with recent experimental data
for undoped La$_2$CuO$_{4}$ compounds.

\end{abstract}

\end{frontmatter}

%

It is well established that spin fluctuations play a prominent role in
high-temperature superconductors. In particular, the parent compounds
display an antiferromagnetic (AF) phase which is usually described as a
two-dimensional (2D) Heisenberg antiferromagnet \cite{kastner}. However,
recent magnetic-susceptibility \cite{Ando} and Raman \cite{Gozar}
experiments in detwinned La$_2$CuO$_4$ (LCO) single crystals can not be
understood within the theory of a conventional antiferromagnet. In
Ref. \cite{noi} we argued that these anomalies originate from the presence,
in the low-temperature orthorhombic (LTO) phase of LCO, of the anisotropic
Dzyaloshinskii-Moriya (DM) and pseudodipolar (XY) interactions between
spins. In this paper we derive explicitly the
long-wavelength limit of the following 2D $S=1/2$ Hamiltonian:
\begin{equation}
H=\sum_{\langle i,j\rangle}J{\bf S}_{i}\cdot{\bf S}_{j}+
{\bf D}_{ij}\cdot\left({\bf S}_{i}\times{\bf
    S}_{j}\right)+{\bf S}_{i}
{\bf \Gamma}_{ij}{\bf S}_{j},
\label{Hamiltonian}
\end{equation}
where ${\bf S}_{i}$ represent the Cu$^{++}$ spins, $J$ is the AF
super-exchange, and ${\bf D}_{ij}$ and ${\bf \Gamma}_{ij}$ are,
respectively, the DM vectors and XY matrices \cite{Shekhtman}.  In the
(xyz) coordinate system of Fig. 1 the DM vectors are oscillating between
neighboring bonds with ${\bf D}_{AB}=(0,d,0)$, and ${\bf D}_{AC}=(d,0,0)$,
while the $\Gamma$ matrices are constant and given by $\Gamma_{AB}=\mathrm
{diag}(\Gamma_1+\Gamma_2, \Gamma_1-\Gamma_2,\Gamma_3)$ and
$\Gamma_{BC}=\mathrm{diag}(\Gamma_1-\Gamma_2,
\Gamma_1+\Gamma_2,\Gamma_3)$. Here $d$ and $\Gamma_{1,2,3}>0$ are of order
$10^{-2}$ and $10^{-4}$, respectively, in units of $J$ \cite{Shekhtman}.
To construct the long wavelength effective theory for the above Hamiltonian
we decompose the unit vector ${\bf \Omega}_i= {\bf S}_i/S$ at site ${\bf
r}_i$ into its slowly-varying staggered and uniform components, ${\bf
\Omega}_i=e^{i{\bf Q}\cdot {\bf r}_i} {\bf n}_i+a{\bf L}_i$, where ${\bf
Q}=(\pi,\pi)$ and $a$ is the lattice parameter.  The constraint ${\bf
\Omega}^2_i=1$ is enforced by ${\bf n}^2_i=1$ and $\bL_i\cdot
\bn_i=0$. Using this decomposition, the Heisenberg part of the Hamiltonian
(\ref{Hamiltonian}) has the standard form \cite{affleck}
\beq
\label{jj}
\LL_{HJ}=J\sum_{\langle i,j\rangle}{\bf S}_{i}\cdot{\bf S}_{j}=
\frac{JS^2}{2}\int \rmd^2 {\bf x} \left[ (\nabla\bn)^2+8\bL^2\right],
\eeq
while the DM and XY terms are given by
\bea 
\LL_{DM}&=&\sum_{\langle i,j\rangle}{\bf D}_{i,j}\cdot\left( {\bf S}_i\times 
{\bf S}_j\right)\nonumber\\
&=&\frac{4S^2}{a}\int \rmd^2 {\bf x} \;\bd_+\cdot(\bn\times \bL)
\label{dd}
\eea
and
\bea
\LL_{XY}&=&\sum_{\langle i,j\rangle}{\bf S}_{i}
{\bf \Gamma}_{ij}{\bf S}_{j}=-\frac{S^2}{a^2}\int\rmd^2 {\bf x} \left[
2\Gamma_+^{\a\b}n_\a n_\b+\right .\nonumber\\
&-&\left.\frac{1}{2}a^2\Gamma_{AB}^{\a\b}(\partial_x n_\a\partial_x n_\b)
-\frac{1}{2}a^2 \Gamma_{BC}^{\a\b}(\partial_y n_\a\partial_y
n_\b)\right].\nonumber\\
\label{gg}
\eea
Observe that only the combination $\bd_+\equiv
(\bD_{AB}+\bD_{AC})/2=(d,d,0)/2$, oriented along the $a$ direction in the
LTO coordinate system, and $\Gamma_+=(\Gamma_{AB}+\Gamma_{AC})/2=
\mathrm{diag}(\Gamma_1,\Gamma_1,\Gamma_3)$ enter the continuum limit of the
model (\ref{Hamiltonian}).  Moreover, since $\Gamma_i\ll J$ we can neglect
the anisotropic corrections in the second line of Eq.\ (\ref{gg}) to the
$(\nabla \bn)^2$ term of Eq.\ (\ref{jj}). Using the fact that $\bn^2=1$ we
can rewrite the first term of Eq.\ (\ref{gg}) as $\LL_{XY}=const+(2S^2/a^2)
\int \rmd^2 {\bf x} (\Gamma_1-\Gamma_3)n_z^2$, which explicitly indicates
that the properties of the model (\ref{Hamiltonian}) only depend on the
difference $\Gamma_1-\Gamma_3>0$. To derive the Euclidean action of the
model (\ref{Hamiltonian}) we introduce the path-integral coherent states
representation of the spin states, which in addition to the previous
contributions gives rise to the (dynamical) Wess-Zumino term \cite{affleck}
\bea
\label{defl}
\LL_{WZ}=\frac{-iS}{a}\int \rmd^2{\bf x}\,\bL\cdot(\bn\times\dot\bn),
\eea
so that the partition function $Z=\int D\bn \delta(\bn^2-1)e^{-{\mathcal
S}}$, with the action ${\mathcal S}=\int \rmd \tau
[\LL_{HJ}+\LL_{DM}+\LL_{XY}+\LL_{WZ}]$.  After integration of the $\bL$
fluctuations we obtain in the LTO coordinate system the following modified
non-linear $\sigma$ model (NLSM) ($\beta=1/T$ and
$\int=\int_{0}^{\beta}\rmd\tau\int\rmd^{2}{\bf x}$):
\begin{equation}
\label{nlsm}
{\mathcal S}=\frac{1}{2gc}\int \left\{(\partial_{\tau}{\bf n})^{2}+
  c^2(\nabla{\bf n})^{2}+{ D_+^2} {n}_a^2+
\Gamma_c\; n_c^2
\right\}.
\end{equation}
The bare coupling constant $g$ and spin velocity $c$ are given by
$gc=8Ja^2$ and $c=2\sqrt{2}JSa$, and we defined $\bD_+=\sqrt{2gcS^2/Ja^2}
\bd_+= (2\sqrt{2}Sd){\bf e}_a$ (${\bf e}_a$ is the unit vector along the
$a$ direction) and $\Gamma_c=(4gc
S^2/a^2)(\Gamma_1-\Gamma_3)=32JS^2(\Gamma_1-\Gamma_3)$.  Eq.\ (\ref{nlsm})
describes a 2D antiferromagnet with an easy axis along the $b$ direction.
Moreover, in the ordered AF phase the system displays also an {\em uniform}
magnetization along the $c$ direction. In fact, the saddle-point value 
for $\bL$, calculated from the Euclidean action $\mathcal S$, is
\beq
\bL=\frac{i}{8aSJ}(\bn\times\dot\bn)+\frac{1}{2Ja}(\bn\times \bd_+).
\label{lsp}
\eeq
Thus, when $\langle \bn \rangle=\sigma_0 {\bf e}_b$ we have $\langle \bL
\rangle = (1/8aSJ)(\sigma_0{\bf e}_b\times D_+{\bf e}_a)=
(\sigma_0D_+/8aSJ){\bf e}_c$. 

We can extend the previous result in Eq.\ (\ref{nlsm}) for the case of 
a multi-layered quasi-2D system, with an exchange coupling $J_\perp$ 
along the $c$ direction, by writing 
${\mathcal S}\rightarrow (1/2gc)\sum_m \int \LL_m$, where 
\bea \LL_m&=&(\partial_{\tau}{\bf n}_m)^{2}+ c^2(\nabla{\bf
n}_m)^{2}+\frac{c^2J_\perp}{a^2J}(\bn_m-\bn_{m+1})^2+ \nonumber\\ &+& {
D_+^m}^2 {n_m}_a^2+ \Gamma_c\; {n_m}_c^2, 
\eea 
where $\bn_m$ is the staggered magnetization in the $mth$ layer. We note
that for a layered system the vector $\bD_+^m=(-1)^mD_+{\bf e}_a$
alternates in sign between neighboring layers, giving rise also to the
staggered order of the uniform $\bL_m$ components, in agreement with the
experiments \cite{Ando,Gozar}.
\begin{figure}[!ht]
\begin{center}
\includegraphics[width=0.4\textwidth]{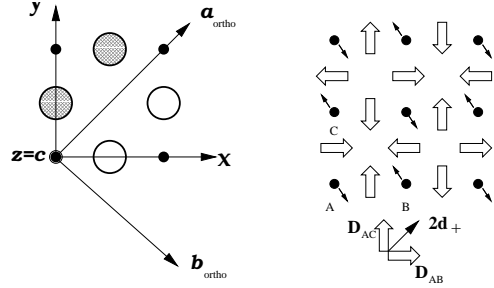}
\end{center}
\caption{Left: tetragonal $(xyz)$ and orthorhombic $(bac)$ reference system
  for the LCO plane. The hatched/empty circles represent the O$^{--}$ ions
  tilted above/below the CuO$_2$ plane. Right: arrangement of the staggered
  magnetization (small black arrows) and DM vectors (open arrows).} 
\label{fig1}
\end{figure}
Finally, in the presence of an uniform magnetic field the total Lagrangian
$\LL=\LL_{HJ}+\LL_{DM}+\LL_{XY}+\LL_{WZ}$ acquires an additional term
$\LL_B=-(S/a)\int d^2{\bf x} {\bf B}\cdot [\bL_m-(\bL_m\cdot\bn_m)\bn_m]$,
where we made explicit the constraint $\bL_m\cdot \bn_m=0$. This leads to
an extra contribution $(1/8aJS)[{\bf B}-({\bf B}\cdot\bn_m)\bn_m]$ to Eq.\
(\ref{lsp}), which after integration over $\bL_m$ has two effects on Eq.\
(\ref{nlsm}): (i) the modification $\partial_{\tau} \bn_m \rightarrow
\partial_{\tau} \bn_m+i{\bf B}\times \bn_m$ of the time derivative
\cite{refB}; (ii) an additional term $(1/gc)\int {\bf B}\cdot({\bf
D}^m_{+}\times{\bf n}_m)$, for each layer. Observe that while the XY
interaction only provides an easy-plane anisotropy for the standard
Heisenberg system, via the $\Gamma_cn_c^2$ term of Eq.\ (\ref{nlsm}), the
DM interaction gives rise to an {\em unconventional} coupling between the
uniform magnetic field and the staggered magnetization. As discussed in
detail in Ref. \cite{noi}, this coupling is responsible for all the
magnetic-susceptibility anomalies observed experimentally \cite{Ando}.


\end{document}